\documentstyle[preprint,aps]{revtex}

\begin{document}

\draft

\title{Gravitational Radiation from Coalescing Binary Neutron Stars}

\author{Xing Zhuge, Joan M. Centrella, and Stephen L. W. McMillan}

\address
{Department of Physics and Atmospheric Science, Drexel University,
Philadelphia, PA 19104}

\maketitle

\begin{abstract}

We calculate the gravitational radiation produced by the merger and
coalescence of inspiraling binary neutron stars using 3-dimensional
numerical simulations.  The stars are modeled as polytropes and start
out in the point-mass limit at wide separation.  The hydrodynamic
integration is performed using smooth particle hydrodynamics (SPH)
with Newtonian gravity, and the gravitational radiation is calculated
using the quadrupole approximation.

We have run several simulations, varying both the neutron star radius
and the equation of state.  The resulting gravitational wave energy
spectra $dE/df$ are rich in information about the hydrodynamics of
merger and coalescence.  In particular, our results demonstrate that
detailed information on both $GM/Rc^2$ and the equation of state can
in principle be extracted from the spectrum.
\end{abstract}


\section{INTRODUCTION}

Coalescing binary neutron stars are among the most promising sources
of gravitational waves for detection by interferometers such as LIGO
and VIRGO \cite{LIGO92,thorne92}.  Recent studies
\cite{how-many} suggest that binary inspiral due to
gravitational radiation reaction, and the eventual coalescence of the
component stars, may be detectable by these instruments at a rate of
several per year. The inspiral phase comprises the last several
thousand binary orbits and covers the frequency range $f\sim
10$--$1000 {\rm Hz}$, where the broad-band interferometers are most
sensitive.  During this stage, the separation of the stars is much
larger than their radii and the gravitational radiation can be
calculated quite accurately using post-Newtonian expansions in the
point-mass limit \cite{post-Newt}.  It is expected that the inspiral
waveform will reveal the masses and spins of the neutron stars, as
well as the orbital parameters of the binary systems
\cite{thorne92,cutler93,CF94}.

When the binary separation is comparable to the neutron star radius,
hydrodynamic effects become dominant and coalescence takes place
within a few orbits.  The coalescence regime probably lies at or
beyond the upper end of the frequency range accessible to broad-band
detectors, but it may be observed using specially designed narrow
band interferometers \cite{narrow} or resonant detectors
\cite{resonant}.  Such observations may yield valuable information
about neutron star radii, and thereby the nuclear equation of state
\cite{cutler93,KLT94,lindblom92}.

Three-dimensional numerical simulations are needed to calculate the
detailed hydrodynamical evolution of the system during coalescence.
Rather than dwell on the uncertain details of the physics of
neutron-star interiors, most studies of this problem have opted
simply to model the neutron stars as polytropes with equation of
state
\begin{equation}
P = K \rho^{\Gamma} = K \rho^{1 + 1/n},
\label{polytrope}  \end{equation}
where $K$ is a constant that measures the specific entropy of the
material and $n$ is the polytropic index.  A choice of $n = 1\
(\Gamma = 2)$ mimics a fairly stiff nuclear equation of
state. Shibata, Nakamura, \& Oohara \cite{SNO,ON} have studied the
behavior of binaries with both synchronously rotating and
non-rotating stars, using an Eulerian code with gravitational
radiation reaction included.  Rasio \& Shapiro \cite{RS92,RS94} have
simulated the coalescence of synchronously rotating neutron-star
binaries using the Lagrangian smooth particle hydrodynamics (SPH)
method with purely Newtonian gravity.  Recently, Davies et
al. \cite{MBD93} have carried out SPH simulations of the inspiral and
coalescence of nonsynchronously rotating neutron stars, focusing on
the thermodynamics and nuclear physics of the coalescence, with
particular application to gamma ray bursts.  All of these studies use
the quadrupole formula to calculate the gravitational radiation
emitted.

Stars in a synchronous binary rotate in the same sense as their
orbital motion, with spin angular velocity equal to the orbital
angular velocity, as seen from a non-rotating frame.  In most close
binary systems (for example, those with normal main-sequence
components) viscosity acts to spin up initially non-rotating stars,
causing them to come into a state of synchronous rotation in a
relatively short period of time \cite{zahn}. However, realistic
neutron star viscosities are expected to be quite small, and recent
work suggests \cite{KBC} that the timescale for synchronization of
neutron star binaries is generally much longer than the timescale for
orbital decay and inspiral due to the emission of gravitational
waves.  Thus neutron star binaries are generally {\it not} expected
to become synchronous as they evolve toward coalescence.

As a complement to full 3-D hydrodynamical simulations, Lai, Rasio,
\& Shapiro \cite{LRS,LRS93} have used quasi-equilibrium methods to
focus on the last 10 or so orbits before the surfaces of the neutron
stars come into contact.  During this time, as tidal effects grow,
the neutron stars are modeled as triaxial ellipsoids inspiraling on a
sequence of quasi-static circular orbits.  Using an approximate
energy variational method these authors have modeled both synchronous
and non-synchronous binaries.  They find that, for sufficiently
incompressible polytropes ($n < 1.2$), the system undergoes a
dynamical instability which can significantly accelerate the secular
orbital inspiral driven by radiation reaction.  (This instability is
driven by Newtonian hydrodynamics; see \cite{KWW93} for the case of
an unstable plunge driven by strong spacetime curvature.)  They have
calculated the evolution of binaries as they approach the stability
limit and the orbital decay changes from secular to dynamical in
character, and have investigated the resulting gravitational wave
emission \cite{LRS}. Their results provide an important component in
understanding the behavior of the full 3-D hydrodynamical models.

We have carried out 3-D simulations of the coalescence of
non-rotating neutron stars using SPH, with particular application to
the resulting gravitational wave energy spectrum $dE/df$.  Our
initial conditions consist of identical spherical polytropes of mass
$M$ and radius $R$ on circular orbits with separations sufficiently
large that tidal effects are negligible.  The stars thus start out
effectively in the point-mass regime.  The gravitational field is
purely Newtonian, with gravitational radiation calculated using the
quadrupole approximation.  To cause the stars to spiral in, we mimic
the effects of gravitational radiation reaction by introducing a
frictional term into the equations of motion to remove orbital energy
at the rate given by the equivalent point-mass inspiral.  As the
neutron stars get closer together the tidal distortions grow and
eventually dominate, and coalescence quickly follows.  The resulting
gravitational waveforms match smoothly onto the point-mass inspiral
waveforms, facilitating analysis in the frequency domain.  We focus
on examining the effects of changing $R$ and the polytropic index $n$
on the gravitational wave energy spectrum $dE/df$.

This paper is organized as follows. In Sec.~\ref{num-tech} we present
a brief description of the numerical techniques we used to do the
simulations.  Sec.~\ref{grav-wave} discusses the calculation of the
gravitational wave quantities, including the spectrum $dE/df$.  The
use of a frictional term in the equations of motion to model the
inspiral by gravitational radiation reaction is discussed in
Sec.~\ref{fric-term}, and the initial conditions are given in
Sec.~\ref{init-cond}. The results of binary inspiral and coalescence
for a standard run with $M = 1.4 {\rm M}_{\odot}$, $R = 10 {\rm km}$,
and polytropic index $n = 1$ ($\Gamma = 2$) are given in
Sec.~\ref{inspiral}, with the frequency analysis and the spectrum
$dE/df$ presented in Sec.~\ref{freq}.  Sec.~\ref{new-param} presents
the results of varying the neutron star radius $R$ and the polytropic
index $n$.  Finally, Sec.~\ref{summary} contains a summary of our
results.

\section{Numerical Techniques}
\label{num-tech}

Lagrangian methods such as SPH \cite{SPH} are especially attractive
 for modeling neutron-star coalescence since the computational
 resources can be concentrated where the mass is located instead of
 being spread over a grid that is mostly empty.  We have used the
 implementation of SPH by Hernquist \& Katz \cite{HK} known as
 TREESPH.  In this code, the fluid is discretized into particles of
 finite extent described by a smoothing kernel.  The use of variable
 smoothing lengths and individual particle timesteps makes the
 program adaptive in both space and time.

Gravitational forces in TREESPH are calculated using a hierarchical
tree method \cite{tree} optimized for vector computers.  In this
method, the particles are first organized into a nested hierarchy of
cells, and the mass multipole moments of each cell up to a fixed
order, usually quadrupole, are calculated.  To compute the
gravitational acceleration, each particle interacts with different
levels of the hierarchy in different ways.  The force due to
neighboring particles is computed by directly summing the two-body
interactions.  The influence of more distant particles is taken into
account by including the multipole expansions of the cells which
satisfy the accuracy criterion at the location of each particle.  In
general, the number of terms in the multipole expansions is small
compared to the number of particles in the corresponding cells.  This
leads to a significant gain in efficiency and allows the use of
larger numbers of particles than would be possible with methods that
simply sum over all possible pairs of particles.

TREESPH uses artificial viscosity to handle the shocks that develop
when stars collide and coalesce.  The code contains three choices for
the artificial viscosity; we have chosen to use the version modified
by the curl of the velocity field.  This prescription reduces the
amount of artificial viscosity used in the presence of curl, and has
proved to be superior to the other options in tests of head-on
collisions of neutron stars \cite{CM} and global rotational
instability \cite{bars}.  Since this has already been discussed in
the literature, we remark only that the artificial viscosity
consists of two terms, one that is linear in the particle velocities
 (with user-specified coefficient $\alpha$) and another
that is quadratic in the velocities
 (with coefficient $\beta$), and refer the interested
reader to references
\cite{HK} and \cite{CM} [see especially their equations (3),
(5), and (6)] for details.

As has been noted above, the neutron stars are not expected to be
synchronously rotating due to their very small physical viscosity.
However in computer simulations numerical viscosity, either present
in the method itself or added explicitly as artificial viscosity, can
have a similar effect and cause the stars to spin up.  We have
monitored this effect in our simulations and have found it to be
small; see Sec.~\ref{inspiral} below.

\section{Calculation of Gravitational Radiation}
\label{grav-wave}

The gravitational radiation in our simulations is calculated in the
quadrupole approximation, which is valid for nearly Newtonian sources
\cite{MTW}. The gravitational waveforms are the
transverse-traceless (TT) components of the metric perturbation
$h_{ij}$,
\begin{equation}
h_{ij}^{TT} = {\frac{G}{c^4}}\, {\frac{2}{r}}
{\skew6\ddot{I\mkern-6.8mu\raise0.3ex\hbox{-}}}_{ij}\,^{TT},
\label{hij-TT}
\end{equation}
where
\begin{equation}
{{I\mkern-6.8mu\raise0.3ex\hbox{-}}}_{ij} = \int\rho \,(x_i x_j -
{\textstyle{\frac{1}{3}}}
\delta_{ij} r^2) \:d^3 r
\label{Iij}
\end{equation}
is the reduced ({\it i.e.} traceless) quadrupole moment of the source
and a dot indicates a time derivative $d/dt$.  Here spatial indices
$i,j=1,2,3$ and the distance to the source $r=(x^2 + y^2 +
z^2)^{1/2}$.  In an orthonormal spherical coordinate system
$(r,\theta,\phi)$ with the center of mass of the source located at
the origin, the TT part of ${I\mkern-6.8mu\raise0.3ex\hbox{-}}_{ij}$
has only four non-vanishing components.  Expressed in terms of
Cartesian components these are ({\it cf.} \cite{KSTC})
\begin{eqnarray}
{I\mkern-6.8mu\raise0.3ex\hbox{-}}_{\theta\theta}&=&
({I\mkern-6.8mu\raise0.3ex\hbox{-}}_{xx}\cos^2\phi+
{I\mkern-6.8mu\raise0.3ex\hbox{-}}_{yy}\sin^2\phi+
{I\mkern-6.8mu\raise0.3ex\hbox{-}}_{xy}\sin 2\phi)\cos^2\theta
\nonumber\\&&\;\;\;\;\;\;\;\;\;\;\;\;\;\;\;+
{I\mkern-6.8mu\raise0.3ex\hbox{-}}_{zz}\sin^2\theta -
({I\mkern-6.8mu\raise0.3ex\hbox{-}}_{xz}\cos\phi+
{I\mkern-6.8mu\raise0.3ex\hbox{-}}_{yz}\sin\phi)\sin 2\theta,
\nonumber\\ {I\mkern-6.8mu\raise0.3ex\hbox{-}}_{\phi\phi}&=&
{I\mkern-6.8mu\raise0.3ex\hbox{-}}_{xx}\sin^2\phi+
{I\mkern-6.8mu\raise0.3ex\hbox{-}}_{yy}\cos^2\phi
-{I\mkern-6.8mu\raise0.3ex\hbox{-}}_{xy}\sin 2\phi, \label{Ispher}\\
{I\mkern-6.8mu\raise0.3ex\hbox{-}}_{\theta\phi}&=&
{I\mkern-6.8mu\raise0.3ex\hbox{-}}_{\phi\theta} \nonumber\\ &=&
\textstyle{\frac{1}{2}}({I\mkern-6.8mu\raise0.3ex\hbox{-}}_{yy}-
{I\mkern-6.8mu\raise0.3ex\hbox{-}}_{xx})\cos\theta\sin
2\phi\nonumber\\&&
\;\;\;\;\;\;\;\;\;\;\;\;\;\;\;+
{I\mkern-6.8mu\raise0.3ex\hbox{-}}_{xy}\cos\theta\cos 2\phi +
({I\mkern-6.8mu\raise0.3ex\hbox{-}}_{xz}\sin\phi -
{I\mkern-6.8mu\raise0.3ex\hbox{-}}_{yz}\cos\phi)\sin\theta.\nonumber
\end{eqnarray}
The wave amplitudes for the two polarizations are then given by
\begin{eqnarray}
h_+&=&\frac{G}{c^4}\frac{1}{r}( {\skew6\ddot{
{I\mkern-6.8mu\raise0.3ex\hbox{-}}}}_{\theta\theta}- {\skew6\ddot{
{I\mkern-6.8mu\raise0.3ex\hbox{-}}}}_{\phi\phi}),
\label{hplus}\\
h_\times &=&\frac{G}{c^4}\frac{2}{r} {\skew6\ddot{
{I\mkern-6.8mu\raise0.3ex\hbox{-}}}}_{\theta\phi}.\label{hcross}
\end{eqnarray}
For an observer located on the axis at $\theta=0, \phi=0$ these
reduce to
\begin{eqnarray}
h_{+{\rm ,axis}}&=&\frac{G}{c^4}\frac{1}{r}( {\skew6\ddot{
{I\mkern-6.8mu\raise0.3ex\hbox{-}}}}_{xx}- {\skew6\ddot{
{I\mkern-6.8mu\raise0.3ex\hbox{-}}}}_{yy}),
\label{hplus-axis}\\
h_{\times{\rm ,axis}} &=&\frac{G}{c^4}\frac{2}{r} {\skew6\ddot{
{I\mkern-6.8mu\raise0.3ex\hbox{-}}}}_{xy}.\label{hcross-axis}
\end{eqnarray}
The angle-averaged waveforms are defined by \cite{KSTC}
\jot 5pt
\begin{eqnarray}
\langle h_+^2 \rangle & = & \frac{1}{4\pi}\int h_+^2 \,
d\Omega \nonumber \\
\langle h_{\times}^2 \rangle & = & \frac{1}{4\pi}
\int h_{\times}^2 \, d\Omega ,
\label{angle-avg}
\end{eqnarray}
which gives
\jot 5pt
\begin{eqnarray}
\frac{c^8}{G^2}
r^2 \langle h_+^2 \rangle & = & {\textstyle{\frac{4}{15}}}
({I\mkern-6.8mu\raise0.3ex\hbox{-}}^{(2)}_{xx}-
{I\mkern-6.8mu\raise0.3ex\hbox{-}}^{(2)}_{zz})^2 +
{\textstyle{\frac{4}{15}}}
({I\mkern-6.8mu\raise0.3ex\hbox{-}}^{(2)}_{yy}-
{I\mkern-6.8mu\raise0.3ex\hbox{-}}^{(2)}_{zz})^2 +
{\textstyle{\frac{1}{10}}}
({I\mkern-6.8mu\raise0.3ex\hbox{-}}^{(2)}_{xx}-
{I\mkern-6.8mu\raise0.3ex\hbox{-}}^{(2)}_{yy})^2 + \nonumber \\ & &
{\textstyle{\frac{14}{15}}}
({I\mkern-6.8mu\raise0.3ex\hbox{-}}^{(2)}_{xy})^2 +
{\textstyle{\frac{4}{15}}}({I\mkern-6.8mu\raise0.3ex\hbox{-}}
^{(2)}_{xz})^2 +
{\textstyle{\frac{4}{15}}}({I\mkern-6.8mu\raise0.3ex\hbox{-}}^
{(2)}_{yz})^2,
\label{h+hx} \\
\frac{c^8}{G^2}
r^2 \langle h_{\times}^2 \rangle & = & {\textstyle{\frac{1}{6}}}
({I\mkern-6.8mu\raise0.3ex\hbox{-}}^{(2)}_{xx}-
{I\mkern-6.8mu\raise0.3ex\hbox{-}}^{(2)}_{yy})^2 +
{\textstyle{\frac{2}{3}}}({I\mkern-6.8mu\raise0.3ex\hbox{-}}
^{(2)}_{xy})^2 +
{\textstyle{\frac{4}{3}}}({I\mkern-6.8mu\raise0.3ex\hbox{-}}
^{(2)}_{xz})^2 +
{\textstyle{\frac{4}{3}}}({I\mkern-6.8mu\raise0.3ex\hbox{-}}
^{(2)}_{yz})^2
. \nonumber
\end{eqnarray}
(Note that equation~(\ref{h+hx}) corrects some typographical errors
in equation (3.12) of \cite{KSTC}.)

The standard definition of gravitational-wave luminosity is
\begin{equation}
L = \frac{dE}{dt} = {\frac15}\frac{G}{c^5} \left\langle \left\langle
{I\mkern-6.8mu\raise0.3ex\hbox{-}}^{(3)}_{ij}
{I\mkern-6.8mu\raise0.3ex\hbox{-}}^{(3)}_{ij} \right \rangle \right
\rangle ,
\label{lum}
\end{equation}
where there is an implied sum on $i$ and $j$, the superscript $(3)$
indicates the third time derivative, and the double angle brackets
indicate an average over several wave periods.  Since such averaging
is not well-defined during coalescence, we simply display the
unaveraged quantity $(G/5c^5)
\textstyle{
{I\mkern-6.8mu\raise0.3ex\hbox{-}}^{(3)}_{ij}
{I\mkern-6.8mu\raise0.3ex\hbox{-}}^{(3)}_{ij}}$ in the plots below.
The energy emitted as gravitational radiation is
\begin{equation}  \Delta E = \int L\; dt .
\label{delta-E}
\end{equation}
The angular momentum lost to gravitational radiation is
\begin{equation}
\frac{dJ_i}{dt} = {\frac{2}{5}}\frac{G}{c^5} \epsilon_{ijk}
\left\langle \left\langle {I\mkern-6.8mu\raise0.3ex\hbox{-}}
^{(2)}_{jm}
{I\mkern-6.8mu\raise0.3ex\hbox{-}}^{(3)}_{km} \right \rangle \right
\rangle\ ,
\label{dJ/dt}
\end{equation}
where $\epsilon_{ijk}$ is the alternating tensor. The total angular
momentum carried away by the waves is
\begin{equation}
\Delta J_i = \int (dJ_i/dt)dt.
\label{delta-J}
\end{equation}
Again, we plot the unaveraged quantity $(2G/5c^5) %
 {\textstyle\epsilon_{ijk}
{I\mkern-6.8mu\raise0.3ex\hbox{-}}^{(2)}_{jm}
{I\mkern-6.8mu\raise0.3ex\hbox{-}}^{(3)}_{km}}$ for $dJ_i/dt$.

The energy emitted in gravitational waves per unit frequency interval
$dE/df$ is given by Thorne \cite{thorne87} in the form
\begin{equation}
\frac{dE}{df} =  \frac{c^3}{G}
 \frac{\pi}{2} (4 \pi r^2) f^2
\langle |\tilde h_+ (f)|^2 + |\tilde h_{\times}(f)|^2 \rangle,
\label{dE/df}  \end{equation}
where $r$ is the distance to the source and the angle brackets denote
an average over all source angles. We define the Fourier transform
$\tilde h(f)$ of any function $h(t)$ by
\begin{equation}
\tilde h(f) \equiv \int_{- \infty}^{+ \infty} h(t) e^{2\pi ift}
 dt
\label{h-FFT} \end{equation}
and
\begin{equation}
h(t) \equiv \int_{- \infty}^{+ \infty} \tilde h(f) e^{-2\pi ift} df.
\label{h-inv-transf} \end{equation}
To calculate the angle-averaged quantity $\langle |\tilde h_+|^2 +
|\tilde h_{\times}|^2 \rangle$ we first take the Fourier transforms
of equations~(\ref{hplus}) and~(\ref{hcross}), to obtain
\begin{eqnarray}
\frac{c^4}{G}
r \tilde h_+&=& {\tilde
 {I\mkern-6.8mu\raise0.3ex\hbox{-}}}^{(2)}_{\theta\theta}- {\tilde
 {I\mkern-6.8mu\raise0.3ex\hbox{-}}}^{(2)}_{\phi\phi} ,
\label{hplusft}\\
 \frac{c^4}{G} r \tilde h_\times &=& 2 {\tilde
{I\mkern-6.8mu\raise0.3ex\hbox{-}}}^{(2)}_{\theta\phi}.
\label{hcrossft}
\end{eqnarray}
The Fourier transforms $\tilde h_+$ and $\tilde h_{\times}$ have the
same angular dependence, given by~(\ref{Ispher}), as $h_+$ and
$h_{\times}$, respectively. The angle averaging
\jot 5pt
\begin{eqnarray}
\langle | \tilde h_+|^2 \rangle
 & = & \frac{1}{4\pi}\int|\tilde h_+|^2 \, d\Omega \nonumber \\
\langle | \tilde h_{\times}|^2 \rangle
& = & \frac{1}{4\pi}
\int| \tilde h_{\times}|^2 \, d\Omega .
\label{angle-avg-ft}
\end{eqnarray}
gives expressions analogous to~(\ref{h+hx}):
\jot 5pt
\begin{eqnarray}
 \frac{c^8}{G^2} r^2 \langle | \tilde h_+|^2 \rangle & = &
 {\textstyle{\frac{4}{15}}} | \tilde
 {I\mkern-6.8mu\raise0.3ex\hbox{-}}^{(2)}_{xx}- \tilde
 {I\mkern-6.8mu\raise0.3ex\hbox{-}}^{(2)}_{zz}|^2 +
 {\textstyle{\frac{4}{15}}} |\tilde
 {I\mkern-6.8mu\raise0.3ex\hbox{-}}^{(2)}_{yy}- \tilde
 {I\mkern-6.8mu\raise0.3ex\hbox{-}}^{(2)}_{zz}|^2 +
 {\textstyle{\frac{1}{10}}}|\tilde
 {I\mkern-6.8mu\raise0.3ex\hbox{-}}^{(2)}_{xx}-
\tilde {I\mkern-6.8mu\raise0.3ex\hbox{-}}^{(2)}_{yy}|^2
  + \nonumber \\
& & {\textstyle{\frac{14}{15}}} |\tilde
{I\mkern-6.8mu\raise0.3ex\hbox{-}}^{(2)}_{xy}|^2 +
{\textstyle{\frac{4}{15}}}|\tilde
{I\mkern-6.8mu\raise0.3ex\hbox{-}}^{(2)}_{xz}|^2 +
{\textstyle{\frac{4}{15}}}|\tilde
{I\mkern-6.8mu\raise0.3ex\hbox{-}}^{(2)}_{yz}|^2,
\label{h+hx-ft} \\
\frac{c^8}{G^2} r^2  \langle | \tilde h_{\times}|^2 \rangle & = &
{\textstyle{\frac{1}{6}}} |\tilde
{I\mkern-6.8mu\raise0.3ex\hbox{-}}^{(2)}_{xx} -\tilde
{I\mkern-6.8mu\raise0.3ex\hbox{-}}^{(2)}_{yy}|^2 +
{\textstyle{\frac{2}{3}}} |\tilde
{I\mkern-6.8mu\raise0.3ex\hbox{-}}^{(2)}_{xy}|^2 +
{\textstyle{\frac{4}{3}}}|\tilde
{I\mkern-6.8mu\raise0.3ex\hbox{-}}^{(2)}_{xz}|^2 +
{\textstyle{\frac{4}{3}}}|\tilde
{I\mkern-6.8mu\raise0.3ex\hbox{-}}^{(2)}_{yz}|^2 .
\nonumber
\end{eqnarray}
We then have
\begin{equation}
\langle |\tilde h_+ (f)|^2 + |\tilde h_{\times}(f)|^2 \rangle
= \langle |\tilde h_+ (f)|^2 \rangle +
\langle |\tilde h_{\times}(f)|^2 \rangle ,
\label{ang-avg}
\end{equation}
which may be substituted into expression~(\ref{dE/df}) for $dE/df$.

We use the techniques of \cite{CM} to calculate the quadrupole moment
and its derivatives.  In particular,
${\skew6\dot{I\mkern-6.8mu\raise0.3ex\hbox{-}}}$ and
${\skew6\ddot{I\mkern-6.8mu\raise0.3ex\hbox{-}}}$ are obtained using
particle positions, velocities, and accelerations already present in
the code to produce very smooth waveforms.  This yields expressions
similar to those of Finn and Evans \cite{FE}.  However,
${I\mkern-6.8mu\raise0.3ex\hbox{-}}^{(3)}_{ij}$ requires the
derivative of the particle accelerations, which is taken numerically
and introduces some numerical noise into $L$ and $dJ_i/dt$.  This
noise can be removed by smoothing; see \cite{CM} for further
discussion.  We have applied this smoothing in producing all graphs
of $L$ and $dJ_i/dt$ in this paper.

\section{Modeling Inspiral by Gravitational Radiation Reaction}
\label{fric-term}

Widely separated binary neutron stars (that is, with separation $a
\gg R$) spiral together due to the effects of energy loss by
gravitational radiation reaction.  Once the two stars are close
enough for tidal distortions to be significant, these effects
dominate and rapid inspiral and coalescence ensue.  In our
calculations we initially place the neutron stars on (nearly)
circular orbits with wide enough separation that tidal distortions
are negligible and the stars are effectively in the point-mass limit.
Since the gravitational field is purely Newtonian and does not take
radiation reaction losses into account, we must explicitly include
these losses to cause inspiral until purely hydrodynamical effects
take over.

To accomplish this, we add a frictional term to the particle
acceleration equations to remove orbital energy at a rate given by
the point-mass inspiral expression (see \cite{MBD93} for a similar
approach).  The gravitational wave luminosity for point-mass inspiral
on circular orbits is \cite{MTW,ST}
\begin{equation}
L_{pm} = \left . \frac{dE}{dt} \right |_{\rm pm} =
\frac{32}{5} \frac{G^4}{c^5} \frac{\mu^2 {\cal M}^3}{a^5} ,
\label{Lpt}
\end{equation}
where $\mu = M_1 M_2/(M_1 + M_2)$ is the reduced mass, ${\cal M} =
M_1 + M_2$ is the total mass of the system, and the subscript ``pm''
refers to point-mass inspiral. For equal mass stars with $M_1 = M_2
\equiv M$ and separation $a = \xi R$, we find
\begin{equation}
\left . \frac{dE}{dt} \right |_{\rm pm} = \frac{64}{5} \frac{c^5}{G}
\frac{1}{\xi^5} \left ( \frac{GM}{c^2R} \right )^5.
\label{dE-pm}  \end{equation}
We then assume that this energy change is due to a frictional force
$\vec f$ that is applied at the center of mass of each star, so that
each point in the star feels the same frictional deceleration.
Dividing the loss equally between the two stars gives
\begin{equation}
\vec f \cdot \vec V = {\frac{1}{2}}
\left . \frac{dE}{dt} \right|_{\rm pm} ,
\end{equation}
where $\vec V$ is the center of mass velocity of the star.  Since
$\vec f$ acts in the direction opposite to $\vec V$ this gives an
acceleration
\begin{equation}
\vec a = \frac{\vec f}{M} = -\frac{1}{2M}
\left . \frac{dE}{dt} \right|_{\rm pm}
\frac{\vec V}{|\vec V|^2}.
\end{equation}
This term is added to the acceleration of every particle, so that
each particle in either star experiences the same frictional
deceleration.  The net effect is that the centers of mass of the
stars follow trajectories that approximate point-mass inspiral.  This
frictional term is applied until tidal effects dominate, leading to
more rapid inspiral and coalescence; see Sec.~\ref{inspiral} and
Fig.~\ref{a-fric-off} below. (Operationally, our assignment of a
particle to a ``star'' is based simply on which body it happened to
belong to initially. Since the frictional term is turned off before
coalescence occurs, the question of what to do after the stars have
merged does not arise.)

The dynamics of polytropes in purely Newtonian gravity is scale free
in the sense that, for a given polytropic index $n$, the results of a
calculation can be scaled for any values of the mass $M$ and radius
$R$.  Inspiral by gravitational radiation reaction introduces the
dimensionless parameter $GM/Rc^2$, as is explicitly evident in
equations~(\ref{dE-pm}) and~(\ref{vx}) for our frictional model of
inspiral.  For neutron stars, $GM/Rc^2$ is determined by the nuclear
equation of state.  In the calculations below, we will vary both $R$
(and hence $GM/Rc^2$) and the polytropic index $n$.

\section{Initial Conditions}
\label{init-cond}

Since our neutron stars start out widely separated with negligible
tidal interaction, they are modeled initially as spherical
polytropes.  Because the timescale for tidal effects to develop is
much greater than the dynamical time $t_D$ for an individual star,
where
\begin{equation}
t_D = \left ( \frac{R^3}{GM} \right ) ^{1/2} ,
\label{tD}
\end{equation}
we start with stable, ``cold'' polytropes produced by the method
discussed in \cite{CM}.  The stars are then placed on a circular
orbit with separation $a_0 = \xi_0 R$ in the center of mass frame of
the system in the $x-y$ plane.  Locating the centers of mass of the
individual stars at $(x,y)$ positions $(\pm a_0/2, 0)$ initially, the
stars are then given the equivalent point-mass velocities for a
circular orbit $V_y = \pm (M/2a_0)^{1/2}$.

To ensure that the stars start out on the correct point-mass inspiral
trajectories, we also give them an initial inward radial velocity
$V_x$ as follows.  For point-mass inspiral the separation $a(t)$ is
given by \cite{MTW}
\begin{equation}
a(t) = a_0 \left ( 1- \frac{t}{\tau_0} \right )^{1/4} ,
\label{a(t)} \end{equation}
where $a_0$ is the separation at the initial time $t=0$ and
\begin{equation}
\tau_0 = \frac{5}{256} \frac{c^5}{G^3} \frac{a_0^4}{\mu {\cal M}^2}
\label{tau0}  \end{equation}
is the inspiral time, {\it i.e.} the time needed to reach separation
$a=0$.  For equal mass stars, the initial inward
velocity is thus
\begin{equation}
V_x = \left . \frac{dr}{dt} \right |_{t=0} =
\left . \frac{1}{2} \frac{da}{dt} \right |_{t=0} =
- \frac{1}{8}
\frac{a_0}{\tau_0} .
\label{vx-1} \end{equation}
Since the stars have initial separation $a_0 = \xi_0
R$ this gives
\begin{equation}
V_x = - \frac{64}{5} \frac{c}{\xi_0^3}
\left ( \frac{GM}{c^2R} \right )^3.
\label{vx}  \end{equation}

The use of the correct initial inspiral trajectory allows us to match
our gravitational waveforms smoothly to the equivalent point-mass
waveforms.  This is important when analyzing the signals in the
frequency domain, as discussed in Sec.~\ref{freq} below.

\section{Binary Inspiral and Coalescence}
\label{inspiral}

We take the case $M = 1.4 {\rm M}_{\odot}$ and $R = 10 {\rm km}$ (so
$GM/Rc^2 = 0.21$), with polytropic index $n = 1$ and initial
separation $a_0 = 4R$ as our standard model, which we refer to as Run
1.  The parameters of this model and the other two models introduced
in Sec.~\ref{new-param} below are presented in
Table~\ref{models-param}. The results of the simulations are
summarized in Table~\ref{models-results}.  Time is measured in units
of the dynamical time $t_D$ given in equation~(\ref{tD}); for Run 1,
$t_D = 7.3 \times 10^{-5} {\rm s}$.

The evolution of this system for the case of $N = 4096$ particles per
star is shown in Fig.~\ref{std-movie}.  Each frame shows the
projection of all particles onto the $x-y$ plane.  As the stars
spiral together their tidal bulges grow.  By $t \sim 100 t_D$ the
stars have come into contact, after which they rapidly merge and
coalesce into a rotating bar-like structure.  Note that the merger is
a fairly gentle process and, in contrast to head-on collisions, does
not generate strong shocks
\cite{RS94,CM,GS}.
Spiral arms form as mass is shed from the ends of the bar.
Gravitational torques cause angular momentum to be transported
outward and lost to the spiral arms.  The arms expand and eventually
form a disk around the central object.  By $t = 200 t_D$ the system
is roughly axisymmetric.

Contour plots at $t = 200 t_D$ reveal more details of the system.  In
Fig.~\ref{contour}(a), which shows a cut along the $x-y$ equatorial
plane, we see that the core is essentially axisymmetric out to
cylindrical radius $\varpi \sim 2R$, where $\varpi = (x^2 +
y^2)^{1/2}$.  As the spiral arms wind up, expand, and merge, the disk
grows increasingly axisymmetric.  In the process the arms expand
supersonically, producing shock heating that causes the disk to puff
up.  This can be seen in Fig.~\ref{contour}(b), which shows density
contours (two per decade in density) for a cut along the meridional
$x-z$ plane.

The angular velocity $\Omega(\varpi)$ of the particles is shown as a
function of cylindrical radius $\varpi$ in Fig.~\ref{omega-std}.  For
our choice of parameters, $\Omega = 1\ (t_D^{-1})$ corresponds to a
spin period $T_{\rm spin} = 0.46 {\rm ms}$.  At $t = 150 t_D$ the
object is in the final stage of its gravitational wave ``ring down''
({\it cf.}  Fig.~\ref{std-waves} below).  Fig.~\ref{omega-std}$a$
shows that the central core $\varpi \lesssim 2R$ is still
differentially rotating at this time (with
$\Omega\sim\varpi^{-0.4}$).  The disk $\varpi \gtrsim 2R$ is also
differentially rotating, with $\Omega\sim\varpi^{-2}$.  By $t = 200
t_D$ the central object has less differential rotation and is more
nearly rigidly rotating, with $\Omega \sim 0.65 t_D^{-1}$, giving a
spin period $T_{\rm spin} \sim 0.71 {\rm ms}$.  The disk is
differentially rotating with $\Omega \sim
\varpi^{-1.7}$. (Recall that Keplerian motion has $\Omega \sim
\varpi^{-1.5}$.)

The mass $m$ contained within cylindrical radius $\varpi$ is plotted
in Fig.~\ref{std-m(varpi)}, showing that $\sim 6\%$ of the mass has
been shed to the disk $\varpi \gtrsim 2R$. Between the times $t = 150
t_D$ and $t = 200 t_D$, some of the matter
in the disk is redistributed out to larger radii.  The specific spin
angular momentum $j(\varpi)$ within cylindrical radius $\varpi$ is
shown in Fig.~\ref{std-j(varpi)}.  About $27 \%$ of the angular
momentum has been shed to the disk, with continued outward transport
of angular momentum within the disk
 taking place between $t=150 t_D$ and $t=200 t_D$.

The gravitational waveforms $r h_+$ and $r h_{\times}$ for an
observer on the axis at $\theta = 0$ and $\varphi = 0$ are shown for
this run in Figs.~\ref{std-waves} (a) and (b), where the solid lines
give the code waveforms and the dashed lines the point-mass results.
For the first couple of orbits after the start of the run ($T_{\rm
orbit} = 2 T_{\rm GW}$) the code waveforms match the point-mass
predictions.  As the tidal bulges grow, the stars spiral in faster
than they would on point-mass trajectories, leading to an increase in
the frequency and amplitude of the gravitational waveforms ({\it
cf.} \cite{LRS}). The gravitational wave amplitudes reach a maximum
during the merger of the two stars at $t \sim 105 - 110 t_D$, then
decrease as the stars coalesce and the spiral arms expand and form
the disk. The peak waveform amplitude $(c^2/GM)rh_+\sim 0.4$
corresponds to a value $h\sim 1.4 \times 10^{-21}$ for a source at
distance r = 20Mpc (the approximate distance to the Virgo
Cluster). By $t\sim 180 t_D$ the gravitational waves have shut off
and the system is essentially axisymmetric.

Fig.~\ref{std-gw} shows (a) the gravitational wave luminosity $L/L_0$
(where $L_0 = c^5/G$) , (b) the energy $\Delta E/Mc^2$ emitted as
gravitational radiation, (c) $dJ_z/dt$ for the gravitational
radiation, and (d) the total angular momentum $\Delta J_z/J$ (where
$J$ is the initial total angular momentum of the system) carried by
the waves.  In all the gravitational-wave quantities, the code
results (solid lines) initially track the point-mass case (dashed
lines) very well, departing significantly from the point-mass
predictions somewhat before the onset of dynamical instability.  The
maximum luminosity is $1.65\times10^{-4}L_0$. This may be compared
with the value of $5.3
\times 10^{-4} L_0$ found by \cite{CM} for the case of a head-on
collision with $GM/Rc^2 = 0.21$; for off-axis collisions on parabolic
orbits, the maximum obtained by those authors was $1.0 \times 10^{-3}
L_0$.  The total energy radiated away after the luminosity departs
 from the point-mass result by more than 10\% is $0.032 Mc^2$.  Again
this can be compared with $0.0025 Mc^2$ for a head-on collision and a
maximum of $0.016 Mc^2$ for off-axis collisions obtained by
\cite{CM}. Although the collisions can achieve a higher gravitational
wave luminosity, they radiate less energy in the form of
gravitational waves overall because they take place on shorter
timescales than the inspiralling binaries.

How sensitive are these results to the resolution of the calculation?
To answer this question we ran the same standard model with different
numbers of particles per star.  Fig.~\ref{std-waves-compare} shows a
comparison of the waveform $r h_+$ for the cases $N = 1024$, $2048$,
and $4096$
 particles per star. It
is clear that the differences in the waveforms are small.  We will
see in Sec.~\ref{freq} below that there are only slight differences
in the frequency domain as well.

In numerical simulations viscosity, whether implicit in the numerical
method or added explicitly as artificial viscosity, can cause
problems by artificially spinning up the stars \cite{MBD93}.  To
monitor this effect in our simulations we calculated the spin angular
momentum of each star about its center of mass and compared this with
the results expected for a synchronously rotating star (the expected
result in the limit of large viscosity).  In general, we have found
that these non-physical viscous effects always remain small in our
simulations.  For example, we ran a test case consisting of initially
non-spinning stars each composed of $N=1024$ particles on a circular
orbit of constant separation $a = 4 R$, with artificial viscosity
coefficients $\alpha = \beta = 0.3$.  After 100 $t_D$ ($\sim 2.8$
orbits), the stars had spin angular momenta $< 2.3\%$ of the
synchronous value. We conclude that numerical and artificial
viscosities play negligible roles in spinning up the stars in our
simulations.

For inspiraling stars, torquing due to the gravitationally-induced
tidal bulges will cause a physical spin-up of the stars \cite{RS94}.
This is shown for the case $N=4096$ particles per star in
Fig.~\ref{spin-4096}, which plots the spin angular momentum of one
star (normalized by the spin of a synchronously rotating star at that
orbital separation) as a function of time.  We see that the spin
angular momentum of the star remains small until contact occurs at $t
\sim 100 t_D$; after this it increases sharply, reaching nearly 70\%
of the synchronous value at $t = 105 t_D$.  (Each ``star'' is
composed of the particles that belonged to it initially, with the
orbital separation of the stars given by the distance between the two
centers of mass.)  Comparison with Fig.~\ref{std-movie} confirms that
this effect is due to the tidal torquing that occurs when the stars
develop large tidal bulges, come into contact, and merge.

Once the stars are close enough for this gravitationally-induced
tidal torquing to be significant, Newtonian gravitational effects
operating on a dynamical timescale dominate the secular radiation
reaction effects, leading to more rapid inspiral, merger, and
coalescence
\cite{LRS}.  We should turn off the gravitational friction term at
some time after the Newtonian tidal torquing takes over and before
the merger occurs, since during the merger the concept of equivalent
point-mass trajectories is meaningless.  We have experimented with
turning off the gravitational friction term at different times and
present the results for our standard run with $N=1024$ particles per
star in Fig.~\ref{a-fric-off}, which shows the center of mass
separation of the two stars as a function of time.
Here, the solid line shows the result of running with the
gravitational friction term left on, and the short dashed lines show
the results of turning this term off at $(1)$ $t=70 t_D$, $(2)$ $t=85
t_D$, and $(3)$ $t=100 t_D$.  The long-dashed line shows the
equivalent point-mass result.

In cases $(1)$ and $(2)$, the stars go into nearly circular orbits
(with eccentricities appropriate to the inspiral radial velocity at
that separation) once the frictional term is turned off.  However,
the trajectory in case $(3)$ is very similar to the result when the
frictional term is left on, indicating that the Newtonian tidal
effects are dominant by this point.  The center of mass separation of
the two stars in this case is $\sim 2.5 R$ at $t=100 t_D$, and then
rapidly decreases.  This result is in good agreement with the
prediction of a dynamical stability limit at $a=2.49 R$ by Lai,
Rasio, and Shapiro \cite{LRS}.  On the basis of these tests, we have
turned off the gravitational friction term at $t = 100 t_D$ for our
standard run, and all the other plots in this paper for this run were
done with this choice.  For each of the runs reported below with
different values of the physical parameters, we have carried out such
experiments to determine the optimal time to turn off the
gravitational friction term, since the time at which the Newtonian
tidal effects dominate differs in each case ({\it cf.} \cite{LRS}).

Finally, we have experimented with the values of the artificial
viscosity coefficients $\alpha$ and $\beta$.  For all runs we used
the values $\alpha = \beta = 0.3$ during the inspiral phase.
However, since shocks occur during the merger, coalescence, and the
expansion of the spiral arms, we ran some tests with different
amounts of artificial viscosity during these regimes.
Fig.~\ref{av-test} shows the waveform $r h_+$ for our standard run
with $N=1024$ particles per star during this phase for three cases:
solid line, $\alpha = \beta = 0.3$; short-dashed line, $\alpha =
0.3$, $\beta = 1$; and long-dashed line, $\alpha = 1$, $\beta = 2$.
Not surprisingly, the amplitude of the waveform is damped as $\alpha$
and $\beta$ are increased.  The low viscosity case conserves energy
to $\sim 2\%$ during the period $100-200$ $t_D$ (after the frictional
term is turned off), indicating that the evolution of the system is
not dominated by strong shocks.  Overall, the differences in energy
conservation for the three cases are not significant.
Therefore, since the low viscosity case
produces the least
damping of the waveform, we chose to use the values $\alpha = \beta =
0.3$ in all of our runs.

\section{Analysis in the Frequency Domain}
\label{freq}

Broad-band detectors such as LIGO and VIRGO should be able to measure
the gravitational waveforms of inspiraling neutron star binaries in
the frequency range $f \sim 10 - 1000 {\rm Hz}$.  Comparison of these
signals with waveform templates derived from post-Newtonian analysis
is expected to yield the neutron-star mass $M$ \cite{cutler93,CF94}.
It is important to develop techniques to measure the neutron-star
radius $R$ since this information, coupled with $M$, can constrain
the equation of state for nuclear matter \cite{lindblom92}.

The actual merger and coalescence stages are driven primarily by
hydrodynamics and are expected to depend on both $R$ and the equation
of state, here parametrized by the polytropic index $n$.  For most
neutron-star binaries, this will take place at frequencies $f > 1000
{\rm Hz}$.  In this regime, shot noise limits the sensitivity of the
broad-band interferometers and so these signals may not be detectable
by them \cite{LIGO92,thorne92}.  However, a set of specially designed
narrow band interferometers \cite{narrow} or resonant detectors
\cite{resonant} may be able to provide information about this high
frequency region \cite{KLT94}.

The merger and coalescence of the neutron stars takes place within
several orbits following initial contact, after which the
gravitational radiation shuts off fairly rapidly as the system
settles into a roughly axisymmetric final state \cite{cutler93}.
This rapid shutoff of gravitational waves is expected to produce a
sharp cutoff in the spectrum $dE/df$.  Since the frequency of the
radiation calculated in the point-mass approximation at separation
$a$ scales as $\sim a^{-3/2}
\sim R^{-3/2}$, a set of narrow-band detectors that can locate the
cutoff frequency where the energy spectrum $dE/df$ drops sharply may
in principle determine the neutron-star radius $R$
\cite{cutler93,KLT94,thorne-pc93}.

We have calculated the spectrum $dE/df$ for our simulations using
equation~(\ref{dE/df}).  For point-mass inspiral, $dE/df \sim
f^{-1/3}$ \cite{thorne87}, where the decrease in energy with
frequency reflects the fact that the binary spends fewer cycles near
a given frequency $f$ as it spirals in.  To see any cutoff frequency
in our data, we need a reasonably long region of point-mass inspiral
in the frequency domain.  Although our runs do start out in the
point-mass regime, the binaries undergo dynamical instability and
rapid merger within just a few orbits.  To compensate for this we
match our waveforms $h_+$ and $h_{\times}$ onto point-mass waveforms
extending back to much larger separations and hence lower
frequencies.

The energy spectrum $dE/df$ for Run 1 with $N=4096$ particles per
star is shown in Fig.~\ref{dE/df-std}.  The solid line shows the
spectrum for the extended waveform including point-mass inspiral, and
the short-dashed line shows the spectrum of the simulation data only.
Note that the two curves match closely.  The separation at which the
data were matched corresponds to frequency $\sim 770 {\rm Hz}$, which
is well within the inspiral regime $dE/df\sim f^{-1/3}$.
Fig.~\ref{dE/df-std} shows that the match is smooth, and does not
affect the merger and coalescence region of the spectrum.  We have
also examined the effect of using different numbers of particles on
$dE/df$ The result is shown in Fig.~\ref{dE/df-std-1024}, which plots
the spectra for Run 1 with $N=1024$ and $4096$ particles per star.
 The use of a
smaller number of particles makes only a slight difference to
$dE/df$.

Examination of Figs \ref{dE/df-std} and \ref{dE/df-std-1024} reveals
several interesting features.  Starting in the point-mass regime, as
$f$ increases, $dE/df$ first drops below the point-mass inspiral
value, reaching a local minimum at $f\sim1500{\rm Hz}$. We identify
this initial dip with the onset of dynamical instability. For the
parameters of Run 1, Lai, Rasio, and Shapiro \cite{LRS} found that
dynamical instability occurs at
separation $a = 2.49 R$; for point-mass inspiral, the frequency
at this separation is
$f_{\rm dyn} = 1566 {\rm Hz}$.
The instability causes the spectrum $dE/df$ to drop below the
point-mass value, since the stars fall together faster than they
would had they remained on strictly point-mass trajectories. For
reference, Fig.~\ref{dE/df-std} also shows the frequency $f_{\rm
contact} = (1/2\pi)(GM/R^3)^{1/2} \sim 2200 {\rm Hz}$,
which is twice the
orbital frequency (in the point-mass limit) at separation $2R$.

This initial fall-off in the spectrum is rather slight. At higher
frequencies, $dE/df$ increases above the point-mass result, reaching
a fairly broad maximum at $f_{\rm peak}\sim2500 {\rm Hz}$, roughly
the frequency of the waves in Fig.~\ref{std-waves} near $t\sim 125
t_D$ (the approximate time at which the gravitational waves shut
off).  To further demonstrate that this feature is associated with
the late-time behavior of the merged system, we have calculated the
spectrum $dE/df$ for the cases in which the waveforms $rh_+$ and
$rh_{\times}$ (including the point-mass inspiral) were truncated at
$t=120 t_D$ and $t=150 t_D$.  The results are shown in
Fig.~\ref{FFT-trunc}, where the solid line shows the spectrum for the
complete waveforms and the dashed lines show the spectra for the
truncated ones.  We see that this peak forms between $t = 120 t_D$
and $t=150 t_D$, and therefore associate it with the transient,
rotating bar-like structure formed immediately following coalescence;
{\it cf.}  Fig.~\ref{std-movie}.  The angular speed of this structure
is $\sim 0.65 t_D^{-1}$ (see Fig.~\ref{omega-std} ), which
corresponds to gravitational radiation with frequencies near $\sim
2800 {\rm Hz}$.  The conclusion that $f_{\rm peak}$ is associated
with a bar is strengthened by Run 3 below, in which the bar survives
for a much longer time and the peak is correspondingly stronger.

Beyond $f_{\rm peak}$, the spectrum drops sharply, eventually rising
again to a secondary maximum at $f_{\rm sec}\sim3200 {\rm Hz}$, too
high to be associated with the bar.  Fig.~\ref{FFT-trunc} shows that
this peak also appears between $t = 120 t_D$ and $t=150 t_D$.  We
attribute this broad high-frequency feature to transient oscillations
induced in the coalescing stars during the merger process---the
result of low-order p-modes with frequencies somewhat higher than the
Kepler frequency in the merging object (see, e.g., \cite{cox80}).

The three frequencies $f_{\rm dyn}$, $f_{\rm peak}$ and $f_{\rm sec}$
serve as a useful means of characterizing our runs. They are
indicated on Fig. \ref{dE/df-std} and are presented in more detail in
Table
\ref{freq-results} below.

\section{The Effects of Changing the Neutron Star Radius and
Equation of State}
\label{new-param}

The energy spectrum $dE/df$ shows rich structure in the frequency
range $f \sim 1000 - 3000 {\rm Hz}$ in which the merger and
coalescence of the neutron stars take place.  To understand how
observations of $dE/df$ might provide information on the neutron star
radius and equation of state, we must investigate the effects of
changing $R$ and the polytropic index $n$.  In this section we
present the results of two runs which begin to explore this parameter
space.  We will continue this study in future papers.

Run 2 is the same as Run 1 except that the initial neutron star
radius is $R = 15 {\rm km}$.  Taking $M = 1.4 {\rm M}_{\odot}$, this
gives $GM/Rc^2 = 0.14$; see Table~\ref{models-param}.  This model was
run with $N = 1024$ particles per star.  The gross features of the
evolution of this model are similar to those found in Run 1.  The
stars first come into contact at $t \sim 250 t_D$.  By the end of the
run, the core $\varpi\lesssim 2R$ is essentially axisymmetric and has
$92 \%$ of the mass and $65 \%$ of the angular momentum.  The disk
extends out to $\sim 10 R$.  The gravitational waveforms $r h_+$ and
$r h_{\times}$ are shown in Figs.~\ref{run9-waves} (a) and (b) for an
observer on the axis at $\theta = 0,\phi = 0$. Fig.~\ref{run9-gw}
shows (a) the gravitational wave luminosity $L/L_0$ and (b) the
energy $\Delta E/Mc^2$ emitted as gravitational radiation.  As in
Fig.~\ref{std-gw}, the time-dependence of the angular momentum
carried away by the waves is quite similar to that of the energy, and
is not presented here.  In these figures, the solid lines give the
code waveforms and the dashed lines the point-mass results.  Some
interesting properties of this model are summarized in
Table~\ref{models-results}.

The energy spectrum $dE/df$ for Run 2 is shown in
Fig.~\ref{dE/df-run9}.  Again, we matched the code data to
point-mass inspiral waveforms for analysis in the frequency domain.
For Run 2 the match occurs at frequency $\sim 420 {\rm Hz}$.
Given the parameters of this run, dynamical instability is expected
to occur at separation $a = 2.49R$ \cite{LRS}; the point-mass
inspiral frequency at this separation is
$f_{\rm dyn} = 852 {\rm Hz}$.
Fig~\ref{dE/df-run9} shows that, as in
Run 1, the spectrum drops below the point-mass inspiral result near
$f_{\rm dyn}$.  The spectrum does not then
rise above the point-mass result at $f_{\rm peak}\sim 1500 {\rm
Hz}$ as in Run 1; however, it does drop sharply just beyond $f_{\rm
peak}$, rising again to a secondary peak at $f_{\rm sec}\sim 1750
{\rm Hz}$.  See Table \ref{freq-results}.

We estimate the frequency of the waves in Fig.~\ref{run9-waves} at
the time when the gravitational radiation shuts off (around $t\sim
270 t_D$) to be $\sim 1300 {\rm Hz}$---that is, close to $f_{\rm
peak}$.  The orbital angular velocity near the end of the run is
$\Omega\sim 0.6 t_D^{-1}$ or $4.5\times 10^3$rad/s; any residual
non-axisymmetric material rotating at this speed would yield
gravitational waves at frequency $\sim 1500 {\rm Hz}$.  Again, we
interpret the secondary peak as the result of high-frequency
oscillations in the merging system. The absence of a strong peak at
$f_{\rm peak}$ and the weaker maximum at $f_{\rm sec}$ is the result
of weaker tidal forces at the point of dynamical instability, leading
to a less pronounced and shorter-lived bar.

Since the frequency of the gravitational radiation for point-mass
inspiral is $\sim a^{-3/2} \sim R^{-3/2}$, we expect the features in
the spectrum for Run 2 to occur at lower frequencies than in Run 1,
roughly in the ratio $f_{\rm Run1}/f_{\rm Run2} \sim (R_{\rm Run
1}/R_{\rm Run 2})^{-3/2} \sim 1.8$. Our numerical simulations do
indeed show this behavior.  For example, the ratio of the frequencies
at which the first peak occurs is $\sim 2500 {\rm Hz} / 1500 {\rm Hz}
\sim 1.7$. The ratio of the frequencies at which the secondary peak
occurs is $\sim 3200 {\rm Hz} / 1750 {\rm Hz} \sim 1.8$.

Run 3 is the same as Run 1 except that we use polytropic index
$n=0.5$ ($\Gamma = 3$).  This model was run with $N = 1024$ particles
per star, with initial separation $a_0 = 4.5 R$.  The stars first
make contact at $t \sim 167 t_D$.  By the end of the run, the core
$\varpi
\lesssim 2R$ has $93\%$ of the mass and $67\%$ of the angular
momentum; the disk extends out to $\sim 50 R$.  The gravitational
waveforms $r h_+$ and $r h_{\times}$ are shown in
Fig.~\ref{run10-waves} for an observer on the axis at $\theta = 0,
\phi = 0$.  Fig.~\ref{run10-gw} shows (a) the gravitational wave
luminosity $L$ and (b) the energy $\Delta E$ emitted as gravitational
radiation.  Again, the solid lines give the code waveforms and the
dashed lines the point-mass results.  Table~\ref{models-results}
summarizes some features of this run.

However, unlike the previous cases, the core of the merged object is
slightly non-axisymmetric, as shown in Fig.~\ref{contour-compare}.
The effect of this rotating, bar-like core can be seen in the
gravitational waveforms $r h_+$ and $r h_{\times}$ in
Fig.~\ref{run10-waves}.  At late times the angular velocity of the
core is $\Omega \sim 0.5 t_D^{-1}$, corresponding to a gravitational
wave frequency $f \sim 2200 {\rm Hz}$.  This agrees with the wave
frequency calculated from Fig.~\ref{run10-waves} at $t \sim 290 t_D$,
confirming that the radiation at late times is due to the rotating
core.  Rasio \& Shapiro \cite{RS94} also found that the coalescence
of a synchronized binary with $n = 0.5$ resulted in a rotating
bar-like core.

The energy spectrum $dE/df$ for Run 3 is shown in
Fig.~\ref{dE/df-run10}.  Here, the match to point-mass inspiral
waveforms occurs at frequency $\sim 640 {\rm Hz}$.
Dynamical instability is expected to occur at separation
$a = 2.76R$ \cite{LRS}, which gives
$f_{\rm dyn} = 1342 {\rm Hz}$.
 Again we see that
the spectrum drops below the point-mass inspiral result near $f_{\rm
dyn}$.  The spectrum then rises to a
sharp peak at $f_{\rm peak}\sim 2200 {\rm Hz}$, drops sharply, then
rises again to a secondary peak at $f_{\rm sec}\sim 2600 {\rm Hz}$.
In this model the gravitational radiation due to the rotating bar is
in the frequency range of the first sharp peak.
See Table \ref{freq-results}.

Lai, Rasio, and Shapiro \cite{LRS} found that the onset of dynamical
instability occurs at separation $a = 2.49 R$ for the parameters of
Run 1 and at $a = 2.76 R$ for the parameters of Run 3.  From this we
estimate that the ratio of frequencies at which the various spectral
features occur should be $f_{\rm Run 1}/f_{\rm Run 3} \sim 1.2$.  Our
simulations approximate this behavior.  For example, the ratio of the
frequencies at which the sharp drop occurs is $\sim 2500 {\rm
Hz}/2200 {\rm Hz} \sim 1.1$.  The ratio of the frequencies at which
the secondary peak occurs is $\sim 3200 {\rm Hz}/2600 {\rm Hz} \sim
1.2$.

\section{Summary and Discussion}
\label{summary}

We have carried out SPH simulations of the merger and coalescence of
identical non-rotating neutron stars modeled as polytropes.  The
stars start out in the point-mass regime and spiral together due to
the effects of gravitational radiation reaction.  Once the stars come
into contact, they rapidly merge and coalesce.  Spiral arms form as
mass is shed from the ends of the central rotating bar-like
structure.  Angular momentum is transported outward by gravitational
torques and lost to the spiral arms.  The arms expand supersonically
and merge, forming a shock-heated axisymmetric disk.  The central
rotating core becomes axisymmetric for $n=1$, with the gravitational
radiation shutting off rapidly after coalescence.  For the stiffer
equation of state $n=0.5$, the rotating core is slightly
non-axisymmetric and considerably longer-lived, and the gravitational
waves decrease more slowly in amplitude.

It is instructive to compare our results with other, related work.
Davies et al. (1993) recently carried out SPH calculations very
similar to ours with $n \sim 0.71$ ($\Gamma = 2.4$).  Their results
for non-rotating stars are similar to ours.  Rasio \& Shapiro
\cite{RS92,RS94} have performed SPH simulations of synchronously
rotating systems.  They found that polytropes with $n=1$ produce an
axisymmetric core, and those with $n=0.5$ yield a non-axisymmetric
core, in agreement with our results.  However, for their
synchronously rotating models, the amplitude of the gravitational
radiation drops off more rapidly after the merger than in our
models. This effect was also seen by Shibata, Nakamura, and Oohara
\cite{SNO} and may be due to the fact that the synchronously rotating
stars are not spinning relative to one another when they merge,
leading to less ``ringing'' of the resulting remnant.

We have also calculated the energy emitted in gravitational waves per
unit frequency interval $dE/df$.  We find that the spectrum gradually
drops below the point-mass inspiral value near the frequency at which
the dynamical instability sets in; this causes the stars to spiral
together faster than they would on point-mass trajectories.  The
spectrum then drops sharply near the frequency at which the waves
 from the main coalescence burst fall off.  Finally, the spectrum
rises again to a secondary peak at larger frequencies, the result of
oscillations that occur during the merger.

The frequencies at which these features in the spectrum occur, as
well as their amplitudes, depend on both the neutron star radius $R$
and the equation of state specified by the polytropic index $n$.  Our
standard model, Run 1, has $R=10 {\rm km}$ and $n=1$.  When we change
just the radius in Run 2 to $R = 15 {\rm km}$, the spectral features
occur at frequencies that are lower by a factor $\sim 1.8$ and the
``gentler'' merger leads to a much lower amplitude in the energy
spectrum, both near $f_{\rm peak}$ and in the secondary maximum.
When instead we change just the polytropic index in Run 3 to $n =
0.5$, the features occur at frequencies that are a factor $\sim 1.2$
lower. The stiffer equation of state results in a longer-lived bar
and a substantially stronger peak amplitude. Measurement of the three
frequencies $f_{\rm dyn}$, $f_{\rm peak}$ and $f_{\rm sec}$, along
with the amplitudes of the spectrum there (relative to the point-mass
result), thus may be used to obtain direct information about the
physical state of the merging neutron stars. While the details of the
peaks depend somewhat on the resolution of our simulations, the
general results described here do not.

The gravitational waveforms and the spectrum $dE/df$ contain much
information about the hydrodynamical merger and coalescence of binary
neutron stars.  Our results show that the characteristic frequencies
depend on both the neutron star radii and the polytropic equation of
state.  We intend to expand our study to include the effects of both
spin and non-equal masses, as well as gravitational radiation
reaction.  In particular, radiation reaction can be expected to
affect the evolution of the rotating bar in Run 3, leading to changes
in the spectrum $dE/df$.  We will present the results of these
studies in future papers.

\acknowledgments
We thank K. Thorne for pointing out the importance of the energy
spectrum $dE/df$ and encouraging this work.  We also thank M. Davies,
D. Kennefick, D. Laurence, and K. Thorne for interesting and helpful
conversations, and L. Hernquist for supplying a copy of TREESPH.
This work was supported in part by NSF grants PHY-9208914 and
AST-9308005, and by NASA grant NAGW-2559.  The numerical simulations
were run at the Pittsburgh Supercomputing Center.

\newpage

\begin{table}[p]
\begin{center}
\begin{tabular}{cccccccc}
Model & $R$& $a_0$ & $GM/Rc^2$ & $t_D$ (ms) & $n$ & $\Gamma$
 & $N$ \\ \tableline
Run 1 & 10 km& $4R$ & 0.21 & 0.073 & 1 & 2 & 4096 \\ Run 2 & 15 km&
$4R$ & 0.14 & 0.13 & 1 & 2 & 1024 \\ Run 3 & 10 km& $4.5R$ & 0.21 &
0.073 & 0.5 & 3 & 1024 \\
\end{tabular}
\end{center}
\caption{Parameters of the models are given. Both the
initial radius $R$ and polytropic index $n$ have been varied. The
neutron star mass is assumed to be $M = 1.4 {\rm M}_{\odot}$ in all
cases.  Run 1 was also run with $N=1024$ and $N=2048$ particles, as
discussed in the text.
\label{models-param}}
\end{table}

\begin{table}[p]
\begin{center}
\begin{tabular}{cccccc}
  Model & $M_{\rm core}$ & $J_{\rm core}$ & $(c^2/GM)rh$ & $L/L_0$ &
 $\Delta E/Mc^2$ \\
\tableline
Run 1 & 94\% & 73\% & 0.4 & $1.65 \times 10^{-4}$ & 0.032 \\ Run 2 &
92\% & 65\% & 0.3 & $2.12 \times 10^{-5}$ & 0.013 \\ Run 3 & 93\% &
67\% & 0.4 & $1.20 \times 10^{-4}$ & $> 0.042$ \\
\end{tabular}
\end{center}
\caption{Results of the simulations are summarized. The core mass
$M_{\rm
core}$ and angular momentum $J_{\rm core}$ refer to material with
cylindrical radius $\varpi~\protect{\lesssim} 2R$.  Peak values of
the waveform $(c^2/GM)rh$ and luminosity $L/L_0$ are given;
$(c^2/GM)rh\sim 0.4$ corresponds to a value $h \sim 1.4 \times
10^{-21}$ at distance $r = 20 {\rm Mpc}$.  The quantity $\Delta
E/Mc^2$ is the energy lost to gravitational radiation after the stars
depart significantly from the point-mass trajectory. It is still
increasing at the end of Run 3 due to the rotating nonaxisymmetric
core.
\label{models-results}}
\end{table}

\begin{table}[p]
\begin{center}
\begin{tabular}{ccccc}
  Model & $a_{\rm dyn}$ &
$f_{\rm dyn}$ (Hz) & $f_{\rm peak}$ (Hz) & $f_{\rm sec}$
(Hz) \\
\tableline
Run 1 & $2.49R$ &
 1566 & 2500 & 3200 \\
Run 2 & $2.49R$ & 852 & 1500 & 1750 \\
Run 3 & $2.76R$ &
1342 & 2200 & 2600 \\
\end{tabular}
\end{center}
\caption{Results of analyzing the simulations in the frequency
domain are summarized.  All frequencies refer to the spectrum $dE/df$;
{\it cf.}  Fig.~\protect{\ref{dE/df-std}}.
  Dynamical instability is expected to occur at
separation $a_{\rm dyn}$; these values are taken from
{\protect{\cite{LRS}}}.
The quantity $f_{\rm
dyn}$ is the gravitational wave frequency for point-mass inspiral
at separation $a_{\rm dyn}$.
\label{freq-results}}
\end{table}

\clearpage

\begin{figure}[p]
\caption{Particle positions are shown projected onto the $x-y$ plane
for Run 1 with $N = 4096$ particles per star.  Here, $M = 1.4
M_{\odot}$, $R = 10 {\rm km}$, polytropic index $n = 1$, and initial
separation $a_0 = 4 R$. The stars are orbiting in the
counterclockwise direction. The vertical axis in each frame is $y/R$
and the horizontal axis is $x/R$.
\label{std-movie}}
\end{figure}

\begin{figure}[p]
\caption{(a) Density contours are shown for a cut
along the $x-y$ plane for Run 1 with $N = 4096$ particles per star at
$t = 200 t_D$.  The contour levels are 0.3, 0.1, 0.03, 0.01,\ldots
(the central density is $0.6 M/R^3$).  (b) The same density contours
as in part (a), but for a cut along the $x-z$ plane.
\label{contour}}
\end{figure}

\begin{figure}[p]
\caption{The angular velocity $\Omega(\varpi)$, where
$\varpi = (x^2 +
y^2)^{1/2}$, is shown for Run 1 with $N=4096$ particles per star.
$\Omega = 1$ ({\it i.e.}~$t_D^{-1}$) corresponds to a spin period
$T_{\rm spin} = 0.46 {\rm ms}$.  (a)\ $t = 150 t_D$, (b)\ $t=200
t_D$.
\label{omega-std}}
\end{figure}

\begin{figure}[p]
\caption{The mass fraction
 $m(\varpi)$ is shown at $t=150 t_D$ and $t=200 t_D$
 for Run 1 with $N=4096$ particles per star.
\label{std-m(varpi)}}
\end{figure}

\begin{figure}[p]
\caption{The specific angular momentum $j(\varpi)$, normalized to
unity for the entire system, is shown at $t=150 t_D$
and $t=200 t_D$ for Run 1 with $N=4096$ particles per
star.
\label{std-j(varpi)}}
\end{figure}

\begin{figure}[p]
\caption{The gravitational waveforms $r h_+$ and $r h_{\times}$ are
shown for an observer on the axis at $\theta = 0$, $\phi = 0$ for Run
1 with $N=4096$ particles per star.  The solid lines give the code
waveforms, and the dashed lines the point-mass results.
\label{std-waves}}
\end{figure}

\begin{figure}[p]
\caption{Various gravitational radiation quantities are shown for Run
1 with $N=4096$ particles per star.  The solid lines show the code
results, and the dashed lines the point-mass values.  (a)
Gravitational wave luminosity $L/L_0$, where $L_0 = c^5/G$; (b)
Energy $\Delta E/Mc^2$ emitted as gravitational radiation; (c)
$dJ_z/dt$; (d) The angular momentum $\Delta J_z/J$ carried away by
the waves, where $J$ is the initial total angular momentum of the
system.
\label{std-gw}}
\end{figure}

\begin{figure}[p]
\caption{A comparison of the waveform $r h_+$ for Run 1 using $N =
1024$, $2048$, and $4096$ particles per star.  The
case $N = 2048$ particles per star was only run for $150 t_D$.
\label{std-waves-compare}}
\end{figure}

\begin{figure}[p]
\caption{The ratio of $J_{\rm spin}$,
the spin angular momentum of a star, to $J_{\rm sync}$, the value for
a synchronously rotating star at that separation, is plotted versus
time for Run 1 with $N=4096$ particles per star. This graph shows
that artificial viscosity does not significantly spin up the star
during the inspiral.  By $t \sim 100 t_D$ the stars are in contact,
and the rapid spin up is due to gravitationally-induced tidal
torquing during the merger process.
\label{spin-4096}}
\end{figure}

\begin{figure}[p]
\caption{The separation between the centers of mass of the stars for
Run 1 with $N = 1024$ particles per star is shown.  The solid line
gives the result of leaving the gravitational friction term on.  The
short-dashed lines show the effect of turning off the friction term
at $(1)$ $t = 70 t_D$, $(2)$ $t = 85 t_D$, and $(3)$ $t = 100 t_D$.
The long-dashed line shows the point-mass result.
\label{a-fric-off}}
\end{figure}

\begin{figure}[p]
\caption{The waveform $r h_+$ is shown for Run 1 with $N = 1024$
particles per star with three different values of the artifical
viscosity coefficients $\alpha$ and $\beta$.
\label{av-test}}
\end{figure}

\begin{figure}[p]
\caption{The gravitational wave energy spectrum $dE/df$ is shown for
Run 1 with $N=4096$ particles per star.  The solid line shows the
result of matching to point-mass inspiral, and the short-dashed line
is the result of using the simulation data only.  The long-dashed
line shows $dE/df$ for point-mass inspiral.  The frequency $f_{\rm
dyn}
= 1566 {\rm Hz}$ is the orbital frequency ($f_{\rm GW}
= 2f_{\rm orb}$) at which dynamical instability occurs
 {\protect{\cite{LRS}}}.  In addition,
$f_{\rm peak} \sim 2500 {\rm Hz}$ and $f_{\rm sec} \sim 3200 {\rm
Hz}$ mark the frequencies of the initial and secondary peaks,
respectively. The frequency $f_{\rm contact}$ corresponding to a
circular point-mass orbit at separation $2R$ is also noted.
\label{dE/df-std}}
\end{figure}

\begin{figure}[p]
\caption{The gravitational wave energy spectrum $dE/df$ is shown for
Run 1 with $N=4096$ particles per star (solid line) and $N=1024$
particles per star (dashed line).
\label{dE/df-std-1024}}
\end{figure}

\begin{figure}[p]
\caption{The gravitational wave energy spectrum $dE/df$ is shown for
Run 1 with $N=4096$ particles per star for the cases in which the
waveforms were truncated at $t=120 t_D$ and $t=150 t_D$.  The solid
lines show the spectrum for the complete waveforms, while the dotted
lines show the truncated cases.
\label{FFT-trunc}}
\end{figure}

\begin{figure}[p]
\caption{
The gravitational waveforms $r h_+$ and $r h_{\times}$ are shown for
an observer on the axis at $\theta = 0$, $\phi = 0$ for Run 2;
compare with Fig.~\protect{\ref{std-waves}}. The solid lines give the
code waveforms, and the dashed lines the point-mass results.
\label{run9-waves}}
\end{figure}

\begin{figure}[p]
\caption{Various gravitational radiation quantities are shown for Run
2.  The solid lines show the code results, and the dashed lines the
point-mass values; compare with Fig.~\protect{\ref{std-gw}}.  (a)
Gravitational wave luminosity $L/L_0$; (b) Energy $\Delta E/Mc^2$
emitted as gravitational radiation.
\label{run9-gw}}
\end{figure}

\begin{figure}[p]
\caption{The gravitational wave energy spectrum $dE/df$ is shown for
Run 2; compare with Fig.~\protect{\ref{dE/df-std}}.
\label{dE/df-run9}}
\end{figure}

\begin{figure}[p]
\caption{The gravitational waveforms $r h_+$ and $r h_{\times}$ are
shown for an observer on the axis at $\theta = 0$, $\phi = 0$ for Run
3; compare with Figs.~\protect{\ref{std-waves}}
and~\protect{\ref{run9-waves}}.  The solid lines give the code
waveforms, and the dashed lines the point-mass results.
\label{run10-waves}}
\end{figure}

\begin{figure}[p]
\caption{Various gravitational radiation quantities are shown for Run
3; compare with Figs.~\protect{\ref{std-gw}}
and~\protect{\ref{run9-gw}}.  The solid lines show the code results,
and the dashed lines the point-mass values.  (a) Gravitational wave
luminosity $L/L_0$; (b) Energy $\Delta E/Mc^2$ emitted as
gravitational radiation.
\label{run10-gw}}
\end{figure}

\begin{figure}[p]
\caption{Density contours are shown for a cut through the
$x-y$ plane of the central regions of (a) Run 1 ($n=1$) and (b) Run 3
($n=0.5$). The same contour levels are plotted in both cases (as in
Fig.~2); however, the central density in (b) is lower than the top
contour in (a).
\label{contour-compare}}
\end{figure}

\begin{figure}[p]
\caption{The gravitational wave energy spectrum $dE/df$ is shown for
Run 3; compare with Figs.~\protect{\ref{dE/df-std}}
and~\protect{\ref{dE/df-run9}}.
\label{dE/df-run10}}
\end{figure}

\newpage
\begin{verbatim}
Compressed PostScript files of Figures 1 - 21 are available
by anonymous ftp from zonker.drexel.edu
in subdirectory papers/ns_coll_1 .
\end{verbatim}

\end{document}